\documentclass[%
reprint,
twocolumn,
 amsmath,amssymb,
 aps,
prb,
]{revtex4-2}

\usepackage{graphicx}
\usepackage[final]{hyperref}
\usepackage{bm}
\usepackage{gensymb}
\usepackage{amsmath}
\usepackage{xcolor}
\usepackage{amssymb}
\usepackage{mathtools}
\usepackage{subcaption}
\usepackage[printonlyused]{acronym}

\definecolor{codegreen}{rgb}{0,0.6,0.1}
\definecolor{codegray}{rgb}{0.5,0.5,0.5}
\definecolor{codepurple}{rgb}{0.58,0,0.82}
\definecolor{backcolour}{rgb}{0.95,0.95,0.92}

\newcommand{\swag}[1]{{\color{codepurple}\ifmmode\text{\footnotesize(SM) #1}\else\footnotesize{(SM) #1}\fi}}
\newcommand{\A}{\bar{\mathbf{A}}}
\newcommand{\W}{\bar{\mathbf{W}}}
\newcommand{\Tr}{\textrm{Tr}}
\newcommand{\x}{\mathbf{x}}
\newcommand{\y}{\mathbf{y}}
\newcommand{\z}{\mathbf{z}}
\newcommand{\1}{\mathbf{1}}
\newcommand{\M}{\mathbf{M}}
\newcommand{\q}{\mathbf{q}}
\newcommand{\R}{\mathbb{R}}
\newcommand{\C}{\mathbf{C}}
\renewcommand{\vec}[1]{\mathbf{#1}}
\newcommand{\set}[1]{\mathbb{#1}}
\newcommand{\cu}{\sigma_i^p}

\makeatletter
\newcommand{\subalign}[1]{%
  \vcenter{%
    \Let@ \restore@math@cr \default@tag
    \baselineskip\fontdimen10 \scriptfont\tw@
    \advance\baselineskip\fontdimen12 \scriptfont\tw@
    \lineskip\thr@@\fontdimen8 \scriptfont\thr@@
    \lineskiplimit\lineskip
    \ialign{\hfil$\m@th\scriptstyle##$&$\m@th\scriptstyle{}##$\hfil\crcr
      #1\crcr
    }%
  }%
}
\makeatother
\begin{document}

\title{Efficient Approximate Methods for  Design of Experiments for Copolymer Engineering}

\author{Swagatam Mukhopadhyay}
\email[]{swag@creyonbio.com}
\affiliation{Creyon Bio, Carlsbad, CA 92010}

\begin{abstract}
We develop a set of algorithms to solve a broad class of Design of Experiment (DoE) problems efficiently. Specifically, we consider problems in which one must choose a subset of polymers to test in experiments such that the learning of the polymeric design rules is optimal. This subset must be selected from a larger set of polymers permissible under arbitrary experimental design constraints. We demonstrate the performance of our algorithms by solving several pragmatic nucleic acid therapeutics engineering scenarios, where limitations in synthesis of chemically diverse nucleic acids or feasibility of measurements in experimental setups appear as constraints. Our approach focuses on identifying optimal experimental designs from a given set of experiments, which is in contrast to traditional, generative DoE methods like \hyperlink{bibd}{BIBD}. Finally, we discuss how these algorithms are broadly applicable to well-established optimal DoE criteria like D-optimality. 

\end{abstract}

\maketitle



\section{Introduction}
\begin{figure*}
    \centering
    \includegraphics[width=\textwidth]{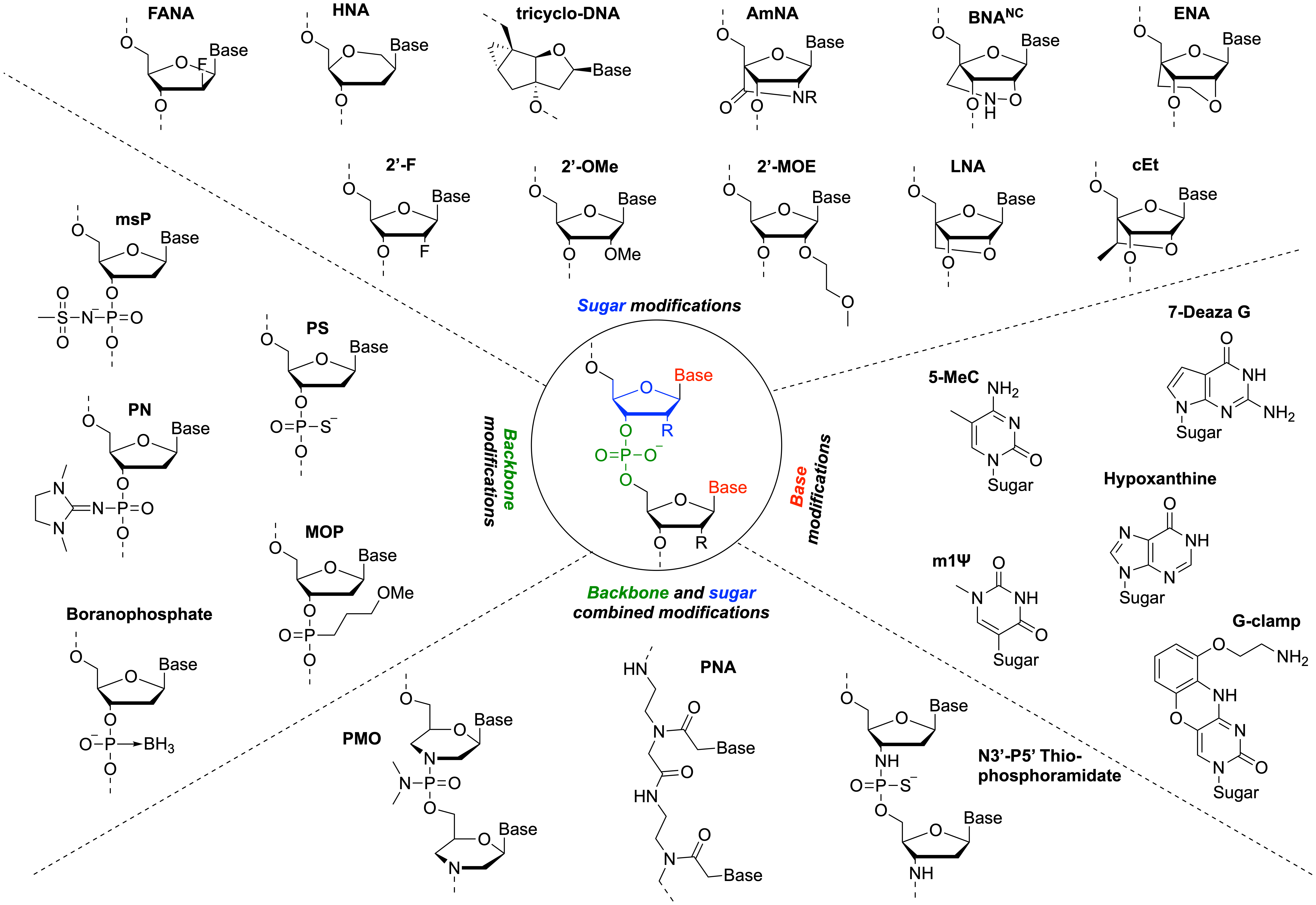}
    \caption{A small fraction of nucleic acid modifications available for constructing synthetic polymeric design of nucleic-acid therapeutics. Center circle shows the prototypical nucleic acids polymer. The polymer consists of a backbone composed of alternating linker (green) and sugar (blue) monomers. The genetic information is encoded in the sequence of bases (red) that are attached to the sugars of the backbone. Naturally occurring nucleic acid polymers most commonly consists of just one type of linker, two types of sugars (Ribose and Deoxyribose), and five types of bases (A,C,T,G, and U). Nucleic acids therapeutics often substitute the naturally occurring monomers for synthetic ones, depicted in the figure, in order to modulate their pharmacological properties.
    }
    \label{fig:nucleic-acids}
\end{figure*}

The problem of engineering polymeric molecules with desirable material properties is ubiquitous in the modern world. A subclass of these problems involves creating chemically diverse polymers (also called copolymers, in contrast to homopolymers) in order to optimize either individual mechanism-of-action related properties or emergent collective properties in solid or liquid phase~\cite{kim2021polymer,martin2023emerging, wu2019machine, sattari2021data}. Such polymeric design questions span a broad swathe of applications---polymeric molecules as therapeutics, for example, \hypertarget{obm}{oligonucleotide-based medicines}---\hyperlink{obm}{OBMs}, like anti-sense oligonucleotides (ASOs), small-interfering RNAs (siRNAs), guide RNAs etc.---polymeric biomaterials~\cite{mcdonald2023applied}, polymeric electrolytes~\cite{li2023machine}, conducting polymers~\cite{hu2024machine}, polymer-based dielectrics~\cite{chen2020frequency}, polymeric delivery systems in bioengineering  (nanocarriers~\cite{ortiz2024machine}, lipid nano-particles~\cite{li2024accelerating, mendes2022nanodelivery}), block copolymers~\cite{meng2014engineering}, etc. These are polymers of relatively fixed length and the challenge is to engineer both their individual molecular and their bulk (bio-)physical,  (bio-)chemical, and material properties, tailored to the application needs. Common problems emerge in the engineering of such polymeric molecules, for example, there are usually design space constraints:
\begin{enumerate}
    \item \emph{Constraints on synthesis} 
    \item \emph{Constraints on experimental feasibility and measurements}
    \item \emph{Constraints on experimental budget}
\end{enumerate}
To illustrate a narrow example of these constraints: \hyperlink{obm}{OBMs} are single or double stranded nucleic acid polymers, used to target RNA or proteins directly, composed of $10-100$ monomers. These molecules are chemically modified for improved bio-stability and favorable pharmacological properties~\cite{wan2016medicinal, egli2023chemistry}. Hundreds of nucleic acid modifications on the linker, sugar or base of these DNA/RNA-like molecules have been reported in the literature~\cite{wan2016medicinal,gait2019advances}, see Fig.~\ref{fig:nucleic-acids} for an illustration. Here, \emph{constraints on synthesis} of \hyperlink{obm}{OBMs} manifests through limitations in commercial availability or explored synthesis routes of the nucleic-acid monomers that combine a choice of linker, sugar, and base modifications. \emph{Constraints on design space} are often biological---the OBMs must be complementary in base sequence to a target loci in a gene. Another biological constraint is the mechanism-of-action the OBM is expected to engage~\cite{gait2019advances, leppek2022combinatorial}. Finally, \emph{constraints in experimental budget} are the enormous costs of pharmacology experiments, preclinical and clinical data generation, etc. For example, animal pharmacology experiments for drug safety can cost in the range of $\$ 10^3 - \$10^6$ (US dollars) per polymer, depending on the stage of the pharmacology study in the highly-regulated drug development process. 

Design of Experiments (DoE) principles can be applied to efficiently uncover the design principles of molecular design, and to build datasets tailored for machine learning---the literature on DoE is voluminous, see Ref.~\cite{lopez2023optimal, stinson2008combinatorial} for pedagogical introductions. For our purposes, we need to consider two related scenarios. The first scenario is when our goal is to create a foundational dataset on polymers, within the aforementioned design constraints, in order to explore the parameter space of design in an unbiased manner. In this scenario, suitable retrospective datasets may not be available. This is the premise of standard DoE paradigms, for example, Balanced Incomplete Block Design~\hypertarget{bibd}{(\hyperlink{bibd}{BIBD})}~\cite{stinson2008combinatorial} etc. Intuitively, such design attempts to explore the parameter space maximally in a minimal set of designs, such that the confounding effects of bias, variability and systematic error are minimized, thereby maximizing the statistical power of the discovered engineering principles. In the second scenario, retrospective datasets exist and perhaps a statistical or machine learning approach reveals certain design principles robustly while also betraying limited statistical power of others---then, what are the minimal set of experiments to optimally learn design rules guided by these initial results? This is the premise of Active Learning, Bayesian Optimization~\cite{garnett2023bayesian} and Bayesian Experimental Design~\cite{lopez2023optimal}. The algorithms presented in this work address aspects of both of these scenarios. 

For polymers, the relevant parameter space are the effect sizes of individual, pairwise and higher-order \hypertarget{kmer}{compositional units} (from here on, simply called \hyperlink{kmer}{k-mers}) in the measurement of interest. Note that \hyperlink{kmer}{k-mers} of importance for the design rules may not be chemical monomers from a synthesis perspective but rather subsequences of the polymer comprising multiple monomers. For nucleic acids, a k-mer is typically composed of $k$ linker-sugar-base monomers. The DoE problem is to perform the minimal number of experiments to identify these \hyperlink{kmer}{k-mers} and quantify their effect sizes for an observable of interest, with low experimental redundancy and high information gain. 

In both of the scenarios discussed, we consider our starting point to be a set of polymers permissible under some constraints. The objective is to find an optimal subset of polymers that fulfil some design rules. In the first half of the paper we explore cases when such design rules can be expressed as pairwise relationships between constituent polymers in the set. This problem maps to finding a densest subgraph of size $K$ in a given graph, which is known to be a NP-hard problem. We present efficient approximate algorithms to solve this---testing performance on graphs that we typically encountered in design of polymers. We note that the graphs we encounter are very distinct compared to typical graphs for which greedy and approximate densest-subgraph algorithms have been reported in the literature~\cite{charikar2000greedy, khuller2009finding, fratkin2006motifcut,boob2020flowless}, which are built on greedy approximations to the \emph{maximum flow} graph algorithm~\cite{goldberg1984finding}, often referred to as the Goldberg's algorithm. All of these algorithms select a subgraph (by using \emph{min-cut}) guided by the degree distribution of the nodes. However, the graphs we encounter are typically uniformly very \emph{dense}--- where the density $d(G)$ of a graph $G$ is defined conventionally as $d(G) = \frac{E}{0.5 \, S(S-1)}$ and $E$ is the number of edges in a graph of $S$ nodes---the densities we encounter range from $0.5-0.95$, see Fig.~\ref{fig:degree_dist}. All nodes are of a degree proportional to the size of the graph $S$, and our problem is to find the $K$-densest subgraph much smaller than graph size, $K \ll S$. The degree distribution of a node belonging to a clique is such a graph is not significantly different compared to a random node typically connected to $50 \% - 95 \%$ of all nodes $S$ in the graph, see Fig.~\ref{fig:degree_dist}. 

A somewhat related past work on finding a $K$-densest subgraph casts the optimization as a maximization problem and random coordinate descent~\cite{sotirov2020solving}. In contrast, as far as we are aware, our approach to finding a $K$-densest subgraph in very dense graphs is novel and we present two efficient algorithms---one which leverages approximate Integer Programming using $l_p$-box constraints~\cite{wu2018ell} and Alternating Direction Method of Multipliers (ADMM)~\cite{boyd2011distributed}, and the other which uses sparsity methods and ADMM~\cite{boyd2011distributed}, see Sections~\ref{sec:AIP}~\ref{sec:Sparse}.

In the second half of the paper, we consider a more general scenario where the features used to encode these polymers are numerical vectors. The objective is to identify a subset of polymers that fulfil a classic design optimality criteria known as \hypertarget{dopt}{\hyperlink{dopt}{D-optimality}}~\cite{lopez2023optimal} which maximizes the determinant of the Fisher Information matrix. This problem is also NP-hard because it involves identifying a sub-matrix of maximal determinant---we present an efficient approximate algorithm and test its performance on pragmatic examples of polymer design. Such optimal design is known to maximize information gain from experiments quantified by the Fisher Information Matrix~\cite{lopez2023optimal}. For linear regression, maximizing Fisher Information is rigorously known to maximize model performance. For nonlinear models, such optimality is usually explored by linearizing around a local region in parameter space~\cite{fedorov2013optimal}. Our solution of \hyperlink{dopt}{D-optimal} design is therefore a powerful tool for refining, in active learning paradigms, both linear and nonlinear regression (ML) models~\cite{fedorov2013optimal}.

\subsection{General considerations in DoE for polymers}
\label{sec:problem_statement}
Polymeric molecules are composed of monomers, and the combination of these monomers are what we control in design space to engineer the target polymeric properties. From a machine learning perspective, these monomers are often not the ideal compositional units to learn and quantify the target polymeric properties---as discussed before, several monomers may comprise the \hyperlink{kmer}{k-mers} most informative and parsimonious in modeling the polymer. In general, these \hyperlink{kmer}{k-mers} may have both independent and correlative contributions (in longer-range interactions between the multiple \hyperlink{kmer}{k-mers} along the polymer). 

For the sake of clarity, we discuss a functional form for fitting observable quantities measured for polymers, without loss of generality of our approach, or suggesting that such a fitting function should be applied in practice. Consider fixed-length polymers composed of \hyperlink{kmer}{k-mers} of fixed size. The design space is the combinatorial chemical space of monomers in each polymer where the monomers are drawn from a fixed library of size $M$---the design problem is to create a minimal set of polymers that explore this chemical space in a \emph{balanced} manner, where in this context \emph{balanced design} means monomers (or monomer pairs, or monomer triplets etc. depending on the design criteria we wish to fulfil) are present in the design the same number of times for every monomer. Second, consider the independent and correlative contributions of k-mers. We define a polymer $\mathcal{P}=\{\cu\vert \cu=1\}$, where each $\cu$ is the indicator variable of a distinct k-mer $i$ in chemical space $\{1,2,\ldots,M\}$ at polymer position $p$. A reasonable probabilistic graphical model to explain a measurement $\vec{Y}$ of the polymer is then:

\begin{align}
\label{eq:general_model}
    \mathcal{P}(\vec{Y}  \vert \{\cu\})  &\sim \exp \left[ - \sum_{ip} {w}_i^p \cu -   \sum_{\subalign{i,&j<i \\p,&q<p}} {w}_{ij}^{pq} \sigma^p_{i}\sigma^q_{j} \right. \nonumber \\
     &\left. - \sum_{\subalign{i, j<&i, k<j \\p, q<&p, r<q}} {w}^{pqr}_{ijk} \sigma^p_{i} \sigma^q_{j} \sigma^r_{k} -  \cdots \right].
\end{align}
In the above equation, $pqr$ are the indices for the position along the polymer, $ijk$ are the indices for the distinct types of the k-mers, and $w$'s are the weights to be learned from data.  The weights $w_{i}^p$ capture the independent contributions of the positional units, $w_{ij}^{pq}$ capture the pairwise contributions of units, and so on. As a first approximation, we may be interested in exploring the independent contributions $\cu$ of possible compositions of units equitably at every position in the polymer. If the positional information is unimportant the design problem simplifies to some extent but does not fundamentally alter the arguments presented below. Similar probabilistic models have been explored in the context of DNA/RNA sequence evolution, see Ref.~\cite{morcos2011direct}. The point is, $\cu, \sigma^p_{i}\sigma^q_{j}$ etc. are physically meaningful features for building ML models for polymers.

Consider a balanced design to explore these independent contributions---what are optimal experiments to learn the weights $w_i^p$, especially, under constraints of permissible polymer designs? Assume we can enumerate all possible designs under some such set of constraints. Say, the $i$ index is over $M$ possible k-mers, the minimal design to optimally explore the \emph{independent} contribution of all the units at every position demands at least $M$ polymers, designed such that every pair of polymers in the set do not share the same $\cu$ at the same position $p$ for all positions. Note that the k-mers could be overlapping along the polymer. This problem maps to finding a dense subgraph problem as follows. Consider a graph with each node as a polymer from the set of possible polymer designs under some constraints. Then the pairwise relationship, where \emph{two polymers do not share the same k-mer $i$ at every position $p$ along the polymer} is a binary relationship. Let fulfilment of this condition create an edge between nodes. Then the minimal set $S$ is the densest subgraph of size $S$ within the graph of the design space (a clique if it exists)---this is a balanced design because every k-mer appears at every position exactly once, considering the count over all the polymers in the set. Across the set $S$, every k-mer $i$ is uniquely present in only one polymer at that position in the set.   

Let us now consider the pairwise contributions. When the positional information of $\cu$ was not important, this problem maps to a classic \hyperlink{bibd}{BIBD} as follows. A $(v, k, \lambda)-$BIBD design is a \emph{block} design where each \emph{block} contains exactly $k$ \emph{points}, and each pair of distinct \emph{points} is contained in exactly $\lambda$ blocks, where the set of \emph{points} are of size $v$. The parameter $v$ is the size of the library $M$ in this case, $k$ is the length of the polymer in k-mers (non-overlapping).  A $(v, k, \lambda)-$BIBD design equitably explores pairwise contribution, every pair is present in exactly $\lambda$ blocks. However, such a \emph{generative} design is restricted to unconstrained cases, and to non-overlapping \hyperlink{kmer}{k-mers}. Again, this problem can be mapped to a dense subgraph problem for $\lambda=1$. The new binary relationship to consider is--\emph{no two polymers in an optimal set share the same pair of \hyperlink{kmer}{k-mers} $i$ and $j$ at the same positions $p$ and $q$, fulfilled for every positions $p$ and $q$ along the polymers.} A large class of practical design problems of polymers map to dense subgraphs. However, the nature of graphs we obtain in such designs is very distinct from social-network graphs, see Section~\ref{sec:results}.

We present a few examples of constraints encountered in the space of nucleic acid design:

\begin{itemize}
    \item \emph{Example 1 :} Identify from the set of all possible single-stranded oligonucleotides (oligos) of length $l$ and set size $S$, all of which have an exact target-loci match to a gene, a minimal subset of size $K$ that is optimally diverse in sequence composition. Arguably, such a design could be useful in unravelling sequence diversity of target loci and oligo sequence needed for mechanism-of-action efficacy.
    \item \emph{Example 2 :} Form a library $S$ of polymers of fixed length $l$ one can chemically synthesize from the library of monomers $M$, where $l \ll M$, identify a minimal set of polymers where all pairwise combinations under synthesis constraints are equitably represented. Arguably, such a design is useful in quantifying the contribution of pairwise monomer compositions
    \item{Example 3: For a set of siRNA designs with constraints in chemical compositions at known positions necessary to retain activity, which target a set of human mRNAs, identify a minimal subset of siRNAs that are maximally diverse in sequence and chemical composition in unconstrained regions of the designed siRNA. Arguably, such a design is optimal in learning pharmacological design rules of such siRNAs }
\end{itemize}

The feature space of polymers may not be limited to individual, pairwise or higher-order correlations, but also include other features--- for example, complementary physical measurements--- in combination. Such features are generally numerical vectors. A more general DoE problem is, given a design matrix $\vec{A}$ of dimension $N \times F$---for a set of $N$ polymers and $F$ dimensional feature space---find a subset $R$ of polymers such that the design is optimal. For example, D-optimal design maximizes the determinant of the Fisher Information matrix $\vec{A}^T \vec{A}$. In Section~\ref{sec:DOE-IP} we present approximate Integer Programming approach to solve \hyperlink{dopt}{D-optimal} for arbitrary design matrices.

\section{Dense Subgraph as an approximate Integer Programming problem}
\label{seq:AIP}
As discussed before, when the constraints can be implemented in pairwise relationships with binary results (\emph{constraint-obeying} and \emph{constraint-violating}) we can map the Design of Experiments problem to finding a dense subgraph in a corresponding relationship graph. In this graph, the nodes are the possible molecules to design. Two nodes are connected by an edge if they obey the set of constraints. Specifically, we want to identify the densest subgraph of size $K$ in a graph of size $S$. We now introduce a fast and approximate dense subgraph algorithm that scales favorably for a large number of nodes and dense connections. In our experience, we observed that most polymeric design problems resulted in very densely connected graphs. 

Denote by $\A$ the adjacency matrix of the complement graph of the graph we want to find the dense subgraph of. Denote by $\x$ a vector of indicator variables of length $S$; $\x = \{x_i \}_{i=1, \dots, S}, x_i  \in \{0, 1\}$, and indicates membership to the dense subgraph. We pose the optimization problem of finding a dense subgraph of size $K$ as follows:
\begin{align}
    & \min_\x  \x^T  \A  \x \ \nonumber \\
    &\textrm{such that } \{x_i \}_{i = 1,\dots, S} \in \{ 0, 1 \} \, \textrm{and} \, \sum_{i=1}^S x_i = K 
\end{align}
Note that the indicator variable $\x$ of the dense subgraph on the original graph would maximize the quadratic term. In order to pose the problem as a minimization, we use the complement graph. 

The above problem is an Integer Programming problem. Also, note that $\A$ is not a positive semi-definite matrix. We next introduce a convex relaxation of the optimization problem, closely following recent work~\cite{wu2018ell}. For the sake of clarity we reproduce the steps here and point out the deviations of this work from~\cite{wu2018ell}. The general idea is to introduce a relaxation of the above Integer Programming problem as an optimization problem amenable to ADMM method (Alternate Direction Method of Multipliers). The problem is not convex, but we show that the approach scales well for large graphs and finds very competitive local solutions. 

\subsection{Approximate Integer Programming: ADMM}
\label{sec:AIP}
Following the work~\cite{wu2018ell}, we relax the integer programming constraint as two simultaneous constraints on real valued $\x$, as a box constraint and a constraint on being within a shifted $l_p$-sphere.  
\begin{align}
    &\{x_i \}_{i = 1,\dots, S} \in \{0, 1 \} \Leftrightarrow \nonumber \\ 
    &\x \in [ 0,1 ]^S \cap \left\{ \x : \Vert \x - \frac{\1}{2} \Vert_p^p = \frac{n}{2^p} \right\}
\end{align}
Intuitively, the points of intersection of the box and the $l_p$-sphere are points where the components $x_i$ are zero or one, the integer programming values. Turning this into a constraint problem on real valued $x_i$ allows us to use ADMM.  

ADMM, originally developed for convex optimization problems, have shown a lot of promise in non-convex optimization. For the sake of clarity and establishing notation throughout the paper, we remind the reader of the ADMM algorithm, see Section 3.1.1 in Ref.~\cite{boyd2011distributed}. The form of optimization problem well-suited to ADMM is:
\begin{align}
\min_{\x, \y} f(\x) + g(\y) \quad \textrm{such that} \quad \M_x  \x  +  \M_y  \y = \q
\end{align}
where $f(\x)$ and $g(\y)$ are typically convex functions. The separability of the objective function and the form of the constraint is what ADMM takes advantage of. Here, $\x \in \R^n$, $\y \in \R^m$, $M_x \in \R^{c \times n}$, $M_y \in \R^{c \times m}$, $\q \in \R^p$, where $c$ is the number of linear constraints. ADMM solves such problems by introducing a penalty parameter, $\rho$, a Lagrange Multiplier $\z$, and introduces the augmented Lagrangian, 
\begin{align}
    & L_\rho(\x, \y, \z) = f(\x) + g(\y) + \z^T  ( \M_x  \x  +  \M_y  \y - \q) + \nonumber \\ 
    & \frac{\rho}{2} \lVert \M_x  \x  +  \M_y  \y - \q \rVert_2^2.
\end{align}
ADMM updates are as follows, see Section 3.1 of ~\cite{boyd2011distributed} for details:
\begin{align}
\x_{k+1} &\coloneq \arg\min_\x\left( f(\x) + \rho/2 \Vert \M_x  \x^k  +  \M_y  \y^k - \q + \z^k \Vert_2^2 \right) \\
\y_{k+1} &\coloneq \arg\min_\y\left( g(\y) + \rho/2 \Vert \M_x  \x^k  +  \M_y  \y^k - \q + \z^k \Vert_2^2 \right) \\
\z^{k+1} &\coloneq \z_k +  \rho(\M_x  \x^{k+1}  +  \M_y  \y^{k+1} - \q) 
\end{align}

In our case, we have the following relaxation of the Integer Programming problem to a real-valued optimization, 
\begin{align}
     & \min_\x  \x^T  \A  \x \ \\
    &\textrm{such that  } 0 \le x_i \le 1 \,  \textrm{and} \, \sum_{i=1}^S x_i = K \, \textrm{and} \, \x \in S_p
\end{align}
Where $S_p$ is the constraint of shifted $l_p$-sphere, $S_p  \coloneq \left\{ \x : \Vert \x - \frac{\1}{2} \Vert_p^p = \frac{S}{2^p} \right\}$. 

We introduce three Lagrange multipliers, corresponding to the three constraints. We write the equality constraint as $\C  \x = \vec{K}$--- here $\C$ is a single row matrix of all ones, the vector $\vec{1}$, and length $S$. For us, $\vec{K}$ is the scalar $K$---the size of the densest subgraph we are searching for. We will take advantage of this single constraint, see later. 

We write the box constraint and the $l_p$ constraints as indicator functions, introducing two variables, $\y_1$ and $\y_2$, with the constraint of $\x = \y_1 = \y_2$, meaning in our ADMM form $\M$'s are positive or negative identity matrices.  Let $g_1(\y_1)$ be the indicator for the box constraint, and $g_2(\y_2)$ the indicator for the $l_p$-sphere constraint $S_p$. The function $g_1(\y_1) = 0 $ if box constraint is obeyed, $S_b \coloneq \left\{ \x : 0 \le x_i  \le 
1 \right\}$, and $g_2(\y_2) = 0$ if $l_p$-sphere constraint $S_p$ is obeyed, where $S_p  \coloneq \left\{ \x : \Vert \x - \frac{\1}{2} \Vert_p^p = \frac{n}{2^p} \right\}$. 

Our ADMM augmented Lagrangian is as follows:
\begin{widetext}
\begin{align}
\label{eq:Lagrangian}
L(\x, \y_1, \y_2, \z_1, \z_2, \z_3) &= \x^T  \A  \x + g_1(\y_1) + g_2(\y_2) + \z_1^T  ( \x - \y_1) + \z_2^T  (\x - \y_2) + \z_3^T  ( \C  \x - \vec{K}) \nonumber  \\
&+ \frac{\rho_1}{2} \Vert \x - \y_1 \Vert_2^2 + \frac{\rho_2}{2} \Vert \x - \y_2 \Vert_2^2  + \frac{\rho_3}{2} \Vert  \C  \x - \vec{K} \Vert_2^2
\end{align}
\end{widetext} 
where the augmented Lagrangian has three penalty parameters $\rho_1, \rho_2, \rho_3$ for the the three constraints, and Lagrange Multipliers $\z_1, \z_2, \z_3$ corresponding to the constraints.  

Now we can obtain the updates for $\x$, the $\y$'s and the $\z$'s. 

\subsubsection{$\x$-update}
We obtain the $x$ update by differentiating Eq.~\ref{eq:Lagrangian} w.r.t. $\x$. We obtain the linear system of equation:
\begin{widetext}
\begin{align}
\label{eq:conjugate_gradient}
    \left( 2 \A + (\rho_1 + \rho_2) \vec{I}  + \rho_3 \C^T \C \right) \x^{k+1} = \rho_1 \y_1^k  + \rho_2 \y_2^k  + \rho_3 \C^T \vec{K}  - \z_1^k - \z_2^k - \C^T  \z_3^k
\end{align}
\end{widetext}
Note that in the above equation, $\C$ is vector, and $\C^T \C$ is a full $ S \times S$ matrix of rank one. The above Eq.~\ref{eq:conjugate_gradient} can be solved by standard Conjugate Gradient methods. However, typically, the adjacency matrix is sparse, and for large graphs being able to use sparse Conjugate Gradient method leads to a substantial speed up. We use the Sherman-Morrison formula to solve this problem using sparse methods alone, by solving two linear equations that are both amenable to sparse Conjugate Gradient solution. We have the following general problem, for a sparse matrix $\M$, a rank-one matrix $\\vec{u}T \vec{u}$ and a vector $\vec{b}$, we need to solve for $\x$:
\begin{align}
    &(\M + \vec{u}^T \vec{u}) \x = \vec{b} \Rightarrow \\
    & \textrm{solve for } \, \M  \x_0 = \vec{b} \textrm{ and } \M  \x_1 = \vec{u} \\
    &\x = \x_0  - \frac{\vec{u}^T  \x_0}{1 + \vec{u}^T  \x_1} \x_1 
\end{align}
In the above equation, $\vec{M}$ is sparse, but $\vec{u}^T \vec{u}$ is not. For us, $\M \coloneq 2 \A + (\rho_1 + \rho_2) \vec{I}$, $\vec{u} \coloneq \sqrt{\rho_3} \C$ and $\vec{b} \coloneq  \rho_1 \y_1  + \rho_2 \y_2  + \rho_3 \C^T \vec{K}  - \z_1 - \z_2 - \C^T  \z_3$. 

In summary, using sparse Conjugate Gradient method we can solve the $\x$-update, deviating from Ref.~\cite{wu2018ell}. 

\subsubsection{$\z$-update}
The updates of all the $\z$-s following ADMM are--
\begin{align}
\z_1^{k+1} &= \z_1^k + \rho_1(\x^{k+1}- \y_1^k) \\
\z_2^{k+1} &= \z_2^k + \rho_2(\x^{k+1}-\y_2^k) \\
\z_3^{k+1} &= \z_3^k + \rho_3(\C^T  \x^{k+1} - \vec{K})  
\end{align}     

\subsubsection{$\y_1$-update}
The indicator function for box constraint is $g(\y_1)$. Derivative of Eq.~\ref{eq:Lagrangian} w.r.t. $\y_1$ provides the update at the $k$-th step, which is an exact solution to the sub-problem:
\begin{align}
    \y_1^{k+1} = \set{P}_{S_b} \left( \x^k + \z_1^k/\rho_1\right)
\end{align}
where  $\set{P}_{S_b}$ is projection to the box constraint $S_b \coloneq 0 < x_i < 1 \,\, \forall i$.

\subsubsection{$\y_2$-update}
The indicator function for the $l_p$-sphere constraint is $g(\y_2)$. Derivative of Eq.~\ref{eq:Lagrangian} w.r.t. $\y_2$ leads to:
\begin{align}
    \y_2^{k+1} = \arg \min_{\y_2 \in S_p} \z_2 ( \x - \y_2) + \frac{\rho_2}{2} \Vert \x - \y_2 \Vert_2^2
\end{align}
Instead of solving this problem exactly, we solve this problem approximately by simply solving the unconstrained problem and projecting on the $l_p$-space. It is known that ADMM can find good solutions even with approximate solution to sub-problems~\cite{boyd2011distributed, wu2018ell}. The approximate solution is:
\begin{align}
    \y_2^{k+1} &= \set{P}_{S_p} \left( \x^{k+1} + \frac{\z_2^{k}}{\rho_2} \right) \\
    \set{P}_{S_p} (\vec{v}) &\coloneq \frac{S^{1/p}}{2} \frac{\left( \vec{v} - \vec{1}/2 \right)}{\Vert \vec{v} - \vec{1}/2 \Vert_p} - \frac{\vec{1}}{2}  
\end{align}
where $\Vert s \Vert_p$ is the $p$-norm, $\vec{1}$ is a vector with all entries of one and size $S$, where $S$ is the size of vector $\x$. 

Following previous work~\cite{wu2018ell} and experimentation, we choose to use $p > 2$, we typically choose $p=3$. We also fix all the penalty parameters $\rho$'s to be equal---$\rho_1 = \rho_2 = \rho_3$. We initialize $\rho_{init} \sim 10$ and increase $\rho$ by a scale factor $\rho_\textrm{scaling} > 1 $, with increasing penalty, after every $t_\textrm{step}$ steps. See~\ref{sec:hyper} for further details. 

\subsubsection{Convergence condition}
Convergence criteria is approximate equality or primary variable $\x$ and dual variables $\y_1$ and $\y_2$ where the relative error in $\x - \y_1$ and $\x - \y_2$ is below a threshold of tolerance, typically, $\max \left(\frac{\Vert \x^k - \y_1 \Vert_2}{\Vert \x^k \Vert_2}, \frac{\Vert \x^k - \y_2 \Vert_2}{\Vert \x^k \Vert_2} \right) < 10^{-5  }$.  

\subsubsection{Optimization considerations}
The above problem in non-convex for two reasons---the $l_p$-sphere constraint is a non-convex constraint and the quadratic terms is not positive semi-definite. This is because the adjacency matrix is not positive semi-definite matrix. We observe that the problem statement of finding $\x$---the indicator variable of the nodes of the dense subgraph---is unchanged by shifting the adjacency matrix to be positive semi-definite by adding a diagonal matrix to $\A$ --- the transformation $\A \rightarrow \A + \vert \lambda_{\min} \vert \vec{I}$ where $\vec{I}$ is the identity matrix, and $\lambda_{min}$ is the smallest (most negative) eigenvalue of $\A$.

We relegate the discussion of results in the combined Results section~\ref{sec:results}.

\subsection{Sparse ADMM: Weighted adjacency matrix and sparse solutions}
\label{sec:Sparse}

In several of design of experiments, the relationship between nodes is not is not binary, meaning, pairwise ``constraint-fulfilling" and ``constraint violating" but is a weighted relationship. Without loss of generality, we assume that small pairwise weights are desired and the problem is to identify the densest set of nodes of size $K$ that minimizes the weights across nodes. For binary relationships, this is equivalent to working with the complement graph in Section~\ref{sec:AIP}, where the densest subgraph was a set of nodes with no edges in the complement graph, and therefore was a minimization problem. To remind the reader that that we continue to work with the complement graph, denoting the weighted adjacency matrix by $\W$, where the bar denotes complement. Without loss of generality, we assume the elements of $\W$ are in the range $[0, 1]$. 

In the weighted context, we seek an optimal fuzzy-indicator variable $\x$ for membership of nodes in the densest subgraph, where the elements $\x_i \in [0,1]$ are real numbers. 

We pose the weighted dense subgraph problem as follows:
\begin{align}
    & \min_\x  \x^T  \W  \x \ +\lambda \vert \x \vert_1 \nonumber \\
    &\ni \{x_i \}_{i = 1,\dots, S} \in [ 0, 1 ] \, \textrm{and} \, \sum_{i=1}^S x_i = K 
\end{align}
where the $l_1$-penalty encourages sparse solution---the membership vector $\x$ is nonzero on a sparse subset of nodes. We interpret $K$ as the sum of total weight on the nonzero nodes. Note that this formulation of the problem maps to the binary solution in Section~\ref{sec:AIP} smoothly by rounding $\W$ to $\A$ and real $\x$ to binary $\x$.

The ADMM augmented Lagrangian for the problem has the same form as Eq.~\ref{eq:Lagrangian} except that now $g(\y_2) = \lambda \vert \y_2 \vert$---the sparsity term, and replace the $l_p$ sphere constraint. The updates for $\x, \z_1, \z_2, \z_3, \y_1$ remain identical. 

The choice of the sparsity parameter is typically $0.5 < \lambda <0.9$. Intuitively, this will drive weights lesser than $0.5$ to zero.

\subsubsection{$\y_2$-update}
Following established methods, see Ref.~\cite{boyd2011distributed}, the $\y_2$ update is a soft threshold function:
\begin{align}
    \y_2^{k+1} = S_{\lambda}\left(\x^k + \frac{\z_2^k}{\rho_2} \right) 
\end{align}
where the soft-threshold operator $S_\lambda$ is applied element-wise to $\x$:
\begin{align}
    S_{\lambda}(x) := \begin{cases}
   x - \lambda & \mbox{if} \, x > \lambda \\
         0 & \mbox{if}\, \vert x \vert \le \lambda \\
          x + \lambda & \mbox{if} \, x < - \lambda,
\end{cases}
\end{align}

\subsection{Results}
\label{sec:results}
We evaluate the algorithm in realistic scenarios of polymer design. Given a set of polymers $\set{S}$ composed of monomer drawn from a monomer library $\set{M}$ of size $M$ and all polymers of equal length $l$, identify a set of polymers that explore the \hyperlink{kmer}{k-mers} of size $k$ equitably, meaning, find a subset such that no two polymers in the subset have the same \hyperlink{kmer}{k-mer} composition at the same position along the polymer. This design principle is relevant in exploring \hyperlink{kmer}{k-mers} of polymers and the chemical diversity of such units. The set $\set{S}$ could be constrained, meaning, exclude some possible monomer combinations.

In our simulation setup, we test our algorithm on randomly generated polymers from a monomer library of size $M = 10$. We consider $k=2$, i.e., overlapping dimers along the polymer. The size of the minimal set $M$ is therefore $M^k = 100$. 

We create $13$ sets of fixed-length polymers, $l$ for lengths $[10, 15, 20, 25, \cdots, 70]$. For each set, we create $900$ random polymers and plant a clique of $100$ polymers, all of identical length. We then test whether we can recover the planted clique using the two algorithms in Section~\ref{sec:AIP} and Section~\ref{sec:Sparse}, and report the distribution of results over $500$ runs for each algorithm. We want to remind the reader that the neither algorithm is a convex optimization problem, and therefore reaching global minima is not guaranteed. The algorithm in Section~\ref{sec:AIP} is non-convex owing to the $l_p$-sphere constraint, and the algorithm in Section~\ref{sec:Sparse} is non-convex owing to the quadratic term not being positive semi-definite. In using this algorithm in the context of binary adjacency matrix of the complement graph, i.e., $\A$, we threshold the continuous solution $\x$ to obtain a binary vector, $x_i > 0.9$ is converted to $1$, and zero otherwise. Surprisingly, this algorithm does better over all compared to Approximate Integer Programming.  

\begin{figure*}
    \includegraphics[width=\textwidth]{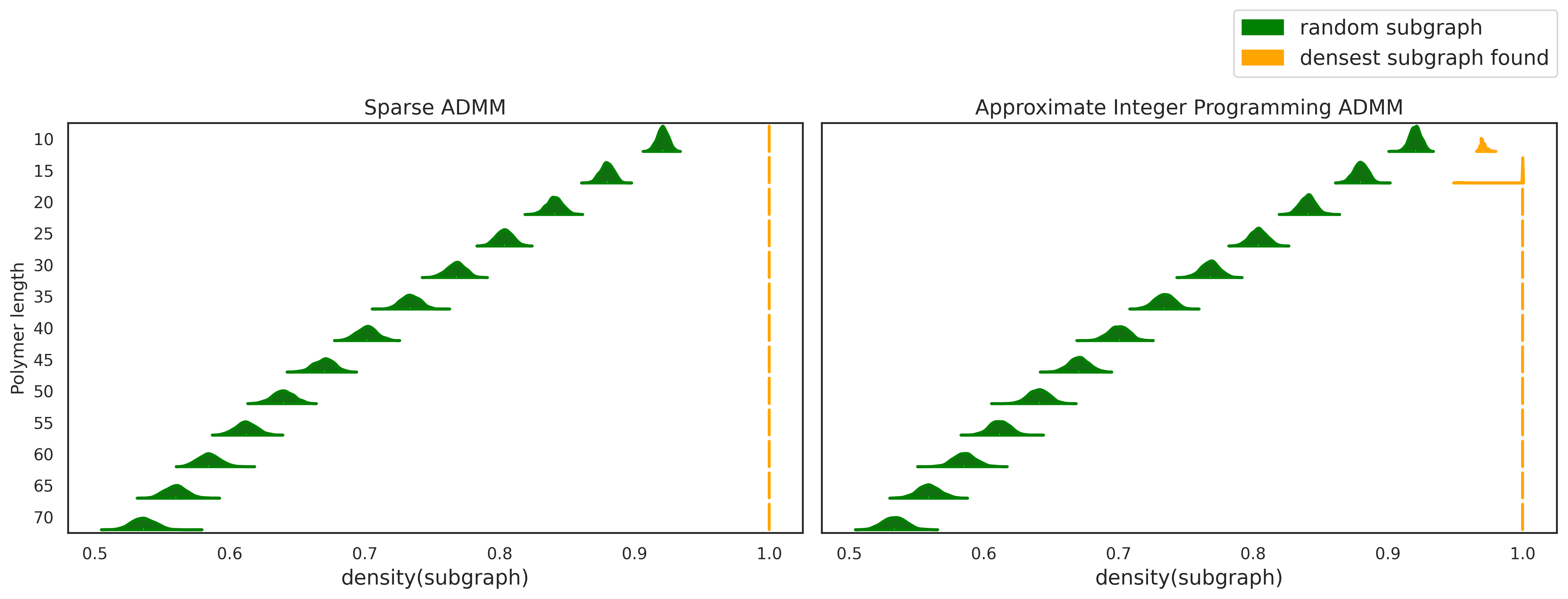}
    \caption{Finding a planted clique in a graph representing a polymer Design of Experiment problem as a function of polymer length, see Section~\ref{sec:results}.
    {\bf Left panel:} Distribution of the density of random subgraphs (green) and subgraphs found by the algorithm of Section~\ref{sec:Sparse} (orange) for polymers of different length. 
    Although our algorithm is not convex, it finds the planted clique (density of one) 100\% of times in $500$ runs. 
    {\bf Right panel:} Same as left, except that the orange density plots correspond to the algorithm of Section~\ref{sec:AIP}. Note that for polymers of length 10 and 15 the algorithm fails to find the planted clique, but it still finds subgraphs that are significantly denser than random subgraphs. This feature indicates that when the algorithm fails to find the optimal solution to the Design of Experiment problem it still performs significantly better than a random solution.}
    \label{fig:results1}
\end{figure*}

In Fig.~\ref{fig:results1} we show comparison of the two algorithms in recovering the hidden clique. Note that the lowest density in our example is roughly $0.5$, where density of a graph $d(G)$ is defined conventionally as $d(G) = \frac{E}{0.5 \, S(S-1)}$ where $E$ is the number of edges in the graph of size $S$. In this example of polymeric design, we are dealing with dense graphs. Notice that over $500$ runs of the Sparse ADMM algorithm (Section~\ref{sec:Sparse}) we recovered the clique in 100\% of runs. For the approximate Integer Programming algorithm, we recover the clique in 100\% of the runs for polymer size $20$ and above. When the algorithm fails to find the clique, it still finds subgraphs that are much denser that random subgraphs. Note that the size of the polymer ($l$) influences the density of the emerging graphs because the likelihood of two polymers not sharing the same \hyperlink{kmer}{k-mer} at the same position, for all positions, is more or less likely as a function of number of positions.

\begin{figure*}
    \centering
    \includegraphics[width=\textwidth]{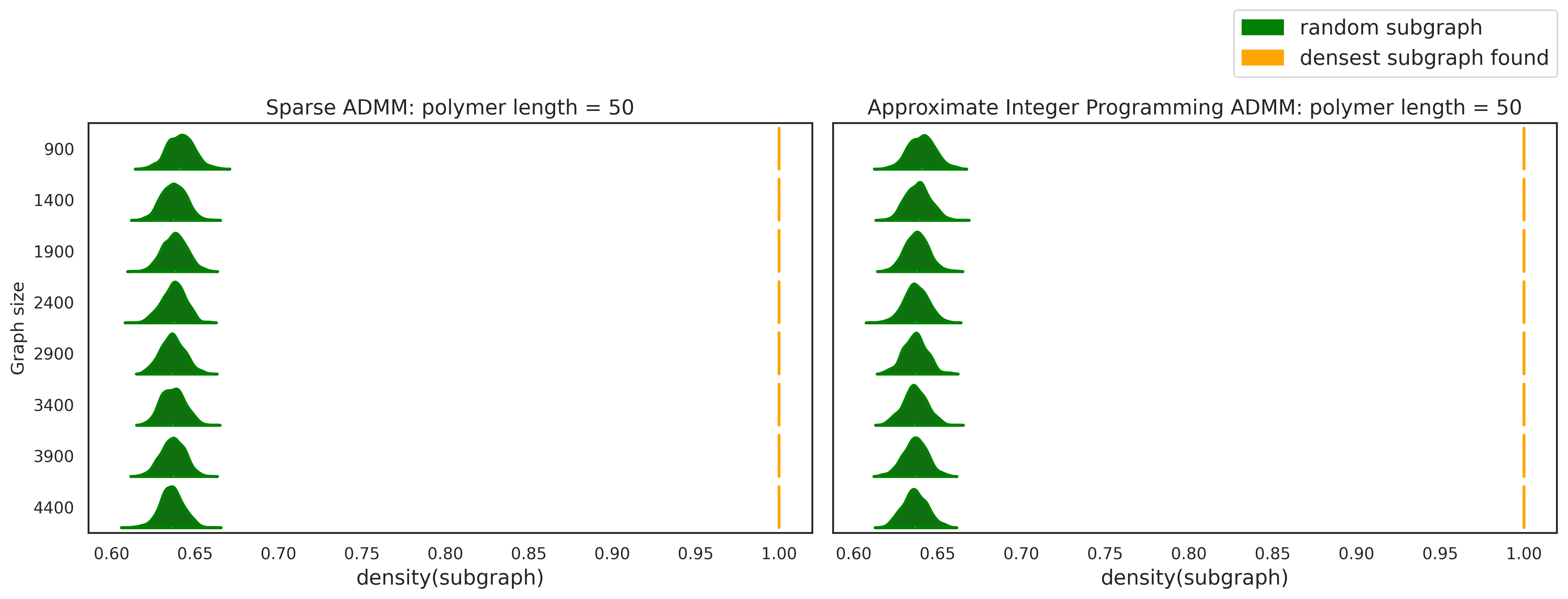}
    \caption{Finding a planted clique in a graph representing a polymer Design of Experiment problem as a function of the graph size. See caption of Fig.~\ref{fig:results1}. Here we vary the size of the random graph with the planted clique and test recovery of the planted clique by the two algorithms for polymer length of $50$. We observe $100\%$ recovery even under such high density of the design graph of $\sim 0.65$, and ratio of planted graph size and design graph size of as low as $1/40$.}
    \label{fig:results2}
\end{figure*}

\begin{figure*}
    \centering
    \includegraphics[width=\textwidth]{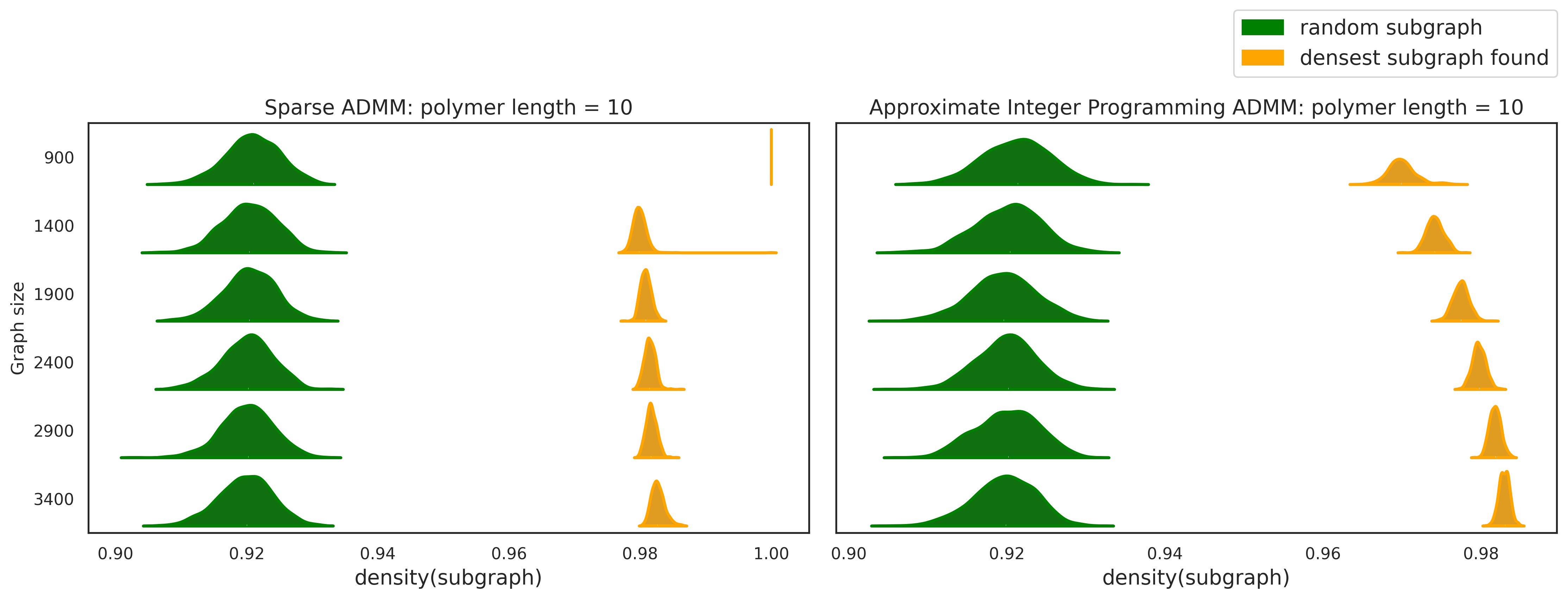}
    \caption{Similar to Fig.~\ref{fig:results2}, except for polymers of length $10$. 
    We observe that for this more difficult problem, where the graph density of the graph is as high as $0.9$,  recovery of the planted clique  (meaning, density of $1$) often fails. However, we recover $K$-densest subgraphs that are still much denser than the random subgraphs. Interestingly, the performance of the two algorithms differ by graph size in this case, showing the value of two different approaches to convex relaxation of a non-convex Integer Programming problem.}
    \label{fig:results3}
\end{figure*}

\begin{figure*}
    \includegraphics[width=0.66\textwidth]{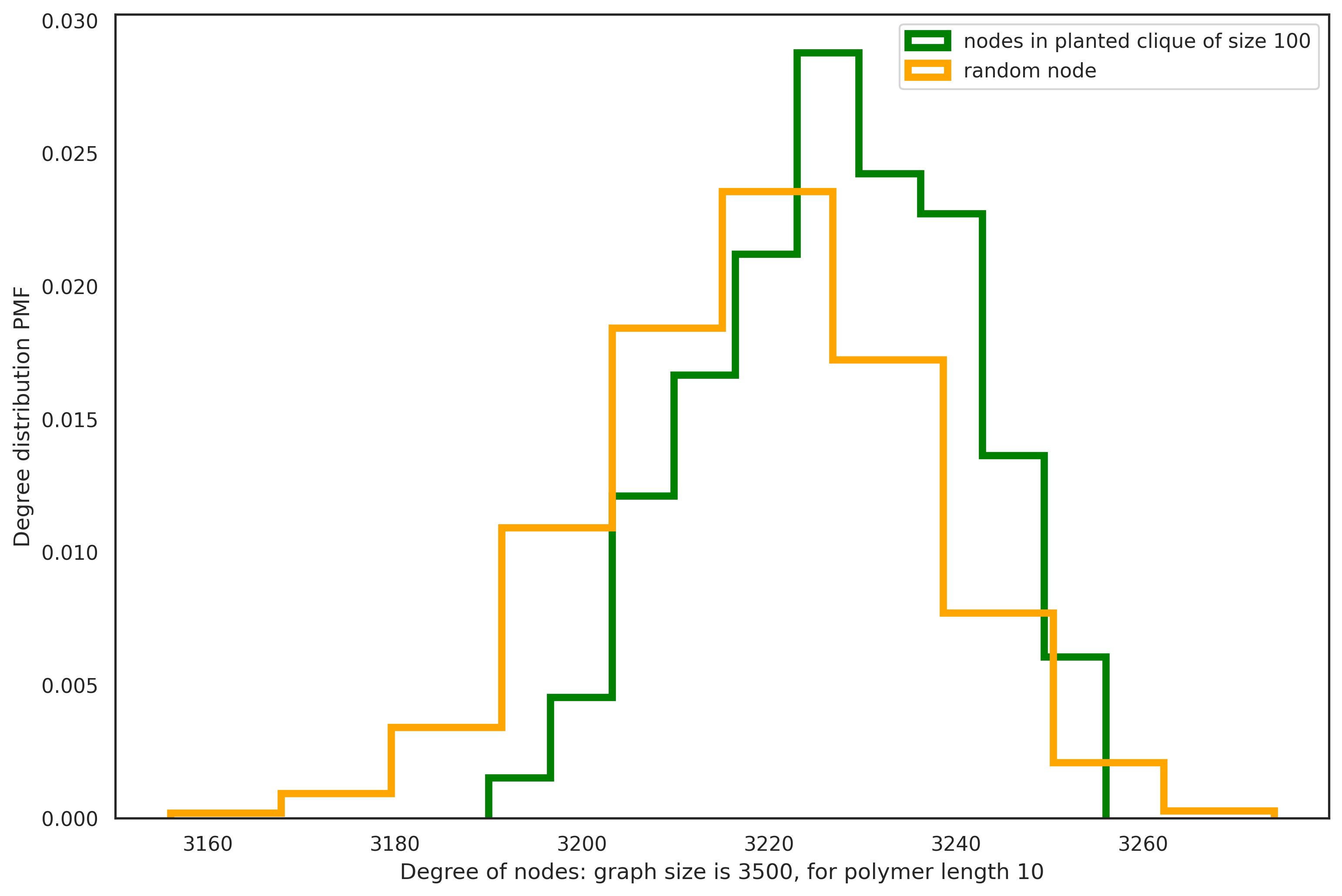}
    \caption{Example degree distribution of the typical graphs we encounter in such DoE. We plot the degree distribution (probability mass function, PMF) for the nodes of the graph of size $3500$ with a $100$ node clique planted, for the polymer of length $10$---see Fig.~\ref{fig:results3}. Note that the degree distribution does not separate the nodes in the clique. The graph is very dense, with median density of $\sim 0.90$.  }
    \label{fig:degree_dist}
\end{figure*}

Next, we study the performance of the algorithm for varying size of the graph. We consider polymer length of $l=50$. We plant a clique in the design (of clique size $100$ as before) and gather statistics on varying sizes of the set of random polymers in the range of $900-3900$---so total graph size in the range of $1000-4000$. Results are shown in Fig.~\ref{fig:results2}. Both algorithms perform very well on these graphs sizes where the ratio of the size of the planted clique to the graph size is $1/40$, with $100 \%$ recovery of planted clique. In Fig.~\ref{fig:results3} we consider polymer length of $l=10$, exploring further the lack of imperfect recovery we observed for that length in Fig.~\ref{fig:results1} for the AIP. We explore the performance across graph sizes, and though the plated clique is not recovered for such dense graph (median density $\sim 0.92$), we still find very dense subgraphs (density $\sim 0.98$). For DoE experiments, this translates to very competitive performance in finding good designs. 

\section{General design matrix---finding an optimal design subset from a given design}
\label{sec:DOE-IP}
In applying ML to polymeric design, the contribution of independent and correlative compositional units (\hyperlink{kmer}{k-mers} or monomers) of the polymer are important, see Eq.~\ref{eq:general_model}. The features of polymers we can directly control in experiments are the presence or absence of these independent, pairwise, etc. compositions. We focus on this feature space of one-hot-encoding. For example, $\cu$ could be a binary vectors of length $M$, encoding presence of unit $i$ at position $p$, $\sigma_{ij}^{pq}$ are binary vectors encoding pairwise presence of units $i,j$ at positions $p,q$ along the polymer, etc. All of these binary features are concatenated to produce the feature vectors, $\x$, say of feature size, $F$. A set of $N$ polymers are featurized as matrix $\vec{A}$  of size $N \times F$. Typically, this is a binary matrix---the results in this section apply to real valued features too. Thus, this is the most general scenario of combinatorial design problem pertaining to polymers.

The design matrix $\vec{A}$, or more specifically, the Fisher Information Matrix $(\vec{A}^T \vec{A})$ has been studied extensively in the literature, where various optimally criteria like A-optimality, \hyperlink{dopt}{D-optimal}, G-optimality, and more generally Schur Optimality, has been discussed~\cite{wallis1988combinatorial}. The problem of selecting a set of rows $\set{R}$ (polymers) from a design matrix $\vec{A}$ with the criteria of the sub-matrix $\vec{A}_R$ being optimal in any of these measures is NP-hard.

Here we present approximate solution to the problem, exploiting the Integer Programming approach above. we focus on \hyperlink{dopt}{D-optimal}, which maximizes the determinant of the information matrix. Which samples (set of rows $\set{R}$) to choose such that the information matrix $\vec{A}_R^T\vec{A}_R$ has largest possible determinant of all possible choices $R$ (of size $R$) can be recast as follows. Consider a vector of weights $\x$. The indicator variable of choosing the row $i$ is $x_i \in [0, 1]$. Denote by $\vec{a}_i$ the $i$-th row of the matrix $\vec{A}$, meaning, the feature vector corresponding to the $i$-th polymer. The problem statement is:
\begin{align}
    &\min_{\x} \left\{ - \log \det \left[ \sum_i x_i \, \vec{a}_i \otimes \vec{a}_i^T \right] \right\}\nonumber \\
    &\ni x_i \in [0, 1] \, \forall \, i = 1, \cdots N, \sum_i x_i = R
\end{align}
We converted a maximization problem into a minimization problem of negative log of the determinant. Note that the Fisher Information Matrix is positive semi-definite. We also express the information matrix over the optimal choice of samples $\set{R}$ as a weighted sum over the outer product of the sample feature vector $\vec{a}_i$, enforcing the sum of weights $x_i$ to be the set size $R$ and the weights $x_i$ being binary indicators for set membership.

We next relax this Integer Programming problem similar to Section~\ref{sec:AIP} by allowing $w_i$ to be real valued and introducing the $l_p$-constraint. The Lagrangian is:

\begin{widetext}
\begin{align}
    \label{eq:DOE-IP}
    L(\x, \y_1, \y_2, \z_1, \z_2, \z_3) &= - \log \det \left[ \sum_i x_i \, \vec{a}_i \otimes \vec{a}_i^T \right] + g_1(\y_1) + g_2(\y_2) + \z_1^T  ( \x - \y_1) + \z_2^T  (\x - \y_2) + \z_3^T  ( \C  \x - \vec{R}) \nonumber  \\
&+ \frac{\rho_1}{2} \Vert \x - \y_1 \Vert_2^2 + \frac{\rho_2}{2} \Vert \x - \y_2 \Vert_2^2  + \frac{\rho_3}{2} \Vert  \C  \x - \vec{R} \Vert_2^2
\end{align}
\end{widetext}
As before, we introduce three Lagrange Multipliers, corresponding to the three constraints. Note that $-\log \det$ of positive-semidefinite matrix is a convex function. Introduce the indicator function, $g(\y)$ for the $l_p$-constraint. We write the equality constraint as $\C  \x = \vec{R}$--- here $\C$ is a single row matrix of all ones, the vector $\vec{1}$, and length $N$. For us, $\vec{R}$ is the scalar $R$---the number of samples in the chosen optimal design. We write the box constraint and the $l_p$ constraints as indicator functions, introducing two variables, $\y_1$ and $\y_2$, with the constraint of $\x = \y_1 = \y_2$. Let $g_1(\y_1)$ be the indicator for the box constraint, and $g_2(\y_2)$ the indicator for the $l_p$-sphere constraint $S_p$. The function $g_1(\y_1) = 0 $ if box constraint is obeyed, $S_b \coloneq \left\{ \x : 0 \le x_i  \le 
1 \right\}$, and $g_2(\y_2) = 0$ if $l_p$-sphere constraint $S_p$ is obeyed, where $S_p \coloneq \left\{ \x : \Vert \x - \frac{\1}{2} \Vert_p^p = \frac{n}{2^p} \right\}$. 

\subsubsection{$\x$-update}
We use the formula, $\frac{\partial }{\partial x}  \log \det \vec{M}= \Tr  \left( \vec{M}^{-1} \frac{\partial \vec{M}}{\partial x} \right)$. 
We obtain, 
\begin{widetext}
\begin{align}
\label{eq:conjugate_gradient2}
    \left( (\rho_1 + \rho_2) \vec{I}  + \rho_3 \C^T \C \right) \x^{k+1}_i = \rho_1 \y_1^k  + \rho_2 \y_2^k  + \rho_3 \C^T \vec{R}  - \z_1^k - \z_2^k - \C^T  \z_3^k + \vec{v}^k
\end{align}
\end{widetext}
 where $v_i^k \coloneq  \Tr \left(\left[ \sum_j^F x_j^k \, \vec{a}_j \otimes \vec{a}_j^T \right]^{-1} \cdot \vec{a}_i \otimes \vec{a}_i^T \right)$ at the $k$-th step. 
 
 In practice, we regularise the matrix and take the inverse of $\sum_j x_j^k \, \vec{a}_j \otimes \vec{a}_j^T 
 + \epsilon \vec{I}$--- where $\epsilon$ is a small diagonal element and $\vec{I}$ is the identity matrix. This matrix inversion naively can be computationally very expensive. We inverse the matrix using the following iteration, using the Sherman-Morrison formula. 
\begin{align}
\vec{S}_j &= \vec{S}_{j-1} - \frac{\vec{w}_j \otimes \vec{w}_j}{1 + \sqrt{x_j}\vec{a}_j \cdot \vec{w}_j}  \\
\vec{w}_j &\coloneq \sqrt{x_j} \vec{S}_{j-1} \vec{a}_j \\
\textrm{where  } \vec{S}_0 &\coloneq \frac{1}{\epsilon} \vec{I} \\
\textrm{providing us  } \vec{S}_F &= \left(\sum_j x_j^k \, \vec{a}_j \otimes \vec{a}_j^T  + \epsilon \vec{I}\right)^{-1} 
\label{eq:costly-inversion}
\end{align}
where the iteration is over $j \in {1 \cdots F}$---the feature dimension. Note that $x_j$ can be negative in intermediate steps-- in computing $\vec{S}_F$ negative components $x_j$ are replaced by zero. 

 In order to further reduce computational cost, observe that when the the update in $\vec{x}$ is small in magnitude, we can compute the matrix inverse in in ~Eq.~\ref{eq:costly-inversion} approximately. We update the matrix, irrespective of the magnitude of change in $\vec{x}$ every $t_{\textrm{v-step}}$. Denoting the deviation $\delta \vec{x}_k \coloneq \vec{x}^k - \vec{x}^{k-1}$. If this deviation is small, we use the approximation, 
 \begin{align}
     \vec{S}_F^k \approx \vec{S}_F^{k-1} - \vec{S}_F^{k-1} \left(\sum_j \delta x_j^k \, \vec{a}_j \otimes \vec{a}_j^T \right) \vec{S}_F^{k-1} 
 \end{align}


The updates for the rest of the variables are nearly identical to Section~\ref{sec:AIP}, with $\vec{R}$ replacing $\vec{K}$ and we skip the steps here. 

\subsubsection{Results}
\label{sec:results-DOPT}
In order to test \hyperlink{dopt}{D-optimal} in realistic scenarios relevant to polymer engineering, we consider the following engineering challenges. Investigation of correlations of features along polymer is often desired---strong position-dependent correlations are common in determining the properties of self-structured polymers like aptamers, or block copolymers, etc.  Consider pairwise \hyperlink{kmer}{k-mers} in positions along the polymer. Concretely, the polymer feature matrix for every polymer is of shape $P \times K$ for $P$ positional pairs $(p, q)$ in consideration, for $K^2$ \hyperlink{kmer}{k-mers} pairs. Note that the design only explore all possible dimer-correlations at the positional pairs $(p,q)$ \emph{independently}. 

A common scenario of such a design is for cases where a ML models on existing data identifies correlations in positional \hyperlink{kmer}{k-mers} to be important, and new experiments are to be designed to optimally investigate such correlations to improve the ML model, or unravel new design principles. We present two simulations. 

We create a set of polymers from a library of $M$ monomers such that all possible (non-overlapping) \hyperlink{kmer}{k-mers} (length $k$) at set of pairs of positions $\{ p_i, p_j\} $ in the polymer are equitably represented, meaning, all pairs appear approximately the same number of times across the designed polymers. This is a D-optimal design in the feature space of pairwise \hyperlink{kmer}{k-mer} features, see below, 

In the test example, we choose $M = 3$, and the  scenario of only two positional pairs, $(p,q),(q, r)$---note that one of the positions $q$ is shared between the two pairwise interactions under investigation.  Note that the design only explore all possible dimer-correlations at $(p,q)$ and $(q,r)$ independently. The minimum number of polymers needed to explore all possible dimer-correlations is $(M^k)^2 = 81$. We create $100$ random polymers and plant the D-optimal design in the random set in order to evaluate the optimal design found by the algorithm. Note that polymer length is not important in this design---the feature space is $P \times  (M^k)^2 = 162$ because there are $P=2$ positional pairs. We wish to find a \hyperlink{dopt}{D-optimal} design of size $(M^k)^2 = 81$. 

Explicitly, the feature space to capture such correlations is as follows. Denote the monomers in the library as $a, b, c$. The possible $16$ dimers are $aa, ab, \cdots cc$. Each polymer is featurized as a  binary matrix $\vec{a}$ of pairwise \hyperlink{kmer}{k-mer} presence/absence at positions $\{(p, q), (q, r)\}$ of each possible dimer---the binary features can be denoted as $(aa_p, aa_q), (aa_q, aa_r) (aa_p, ab_q), (aa_q, aa_r)$ $\cdots$ $ (cc_p, cc_q), (cc_q, cc_r)$, where subscript on dimer implies position of that dimer on the polymer etc. The objective is to find approximate D-optimal design from a set of designs given to us---we evaluate \hyperlink{dopt}{D-optimal} by how large the sum of log eigenvalues of the information matrix $\vec{A}^T_R \vec{A}_R$ is compared to random subset and the planted \hyperlink{dopt}{D-optimal} design.

It is easy to see that the $\vec{A}^T_R \vec{A}_R$ for the planted D-optimal design is a diagonal square matrix of size $81$ with all diagonal elements equal to $2$.  

\begin{figure*}
    \centering
    \includegraphics[height=3 in]{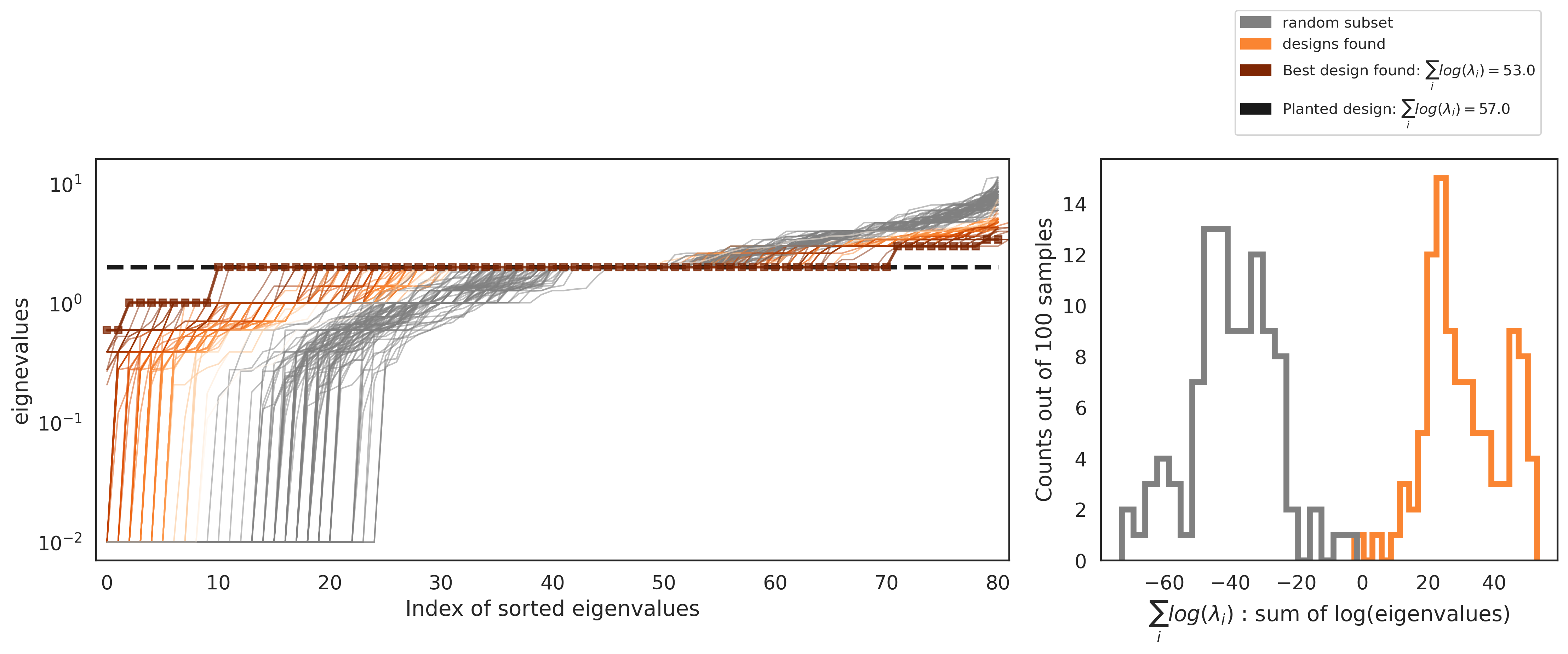}
    \caption{Test of DoE Integer Programming algorithm described in Section~\ref{sec:DOE-IP}. In this simulation, a design of pairwise positional k-mers in a polymer at two positions along the polymer was explored, see main text. The design problem is to identify $81$ polymers from a set of $181$--- where $100$ polymers are random, and $81$ are planted D-optimal design. {\bf Left panel:} Sorted eigenvalues of regularized Fisher Information Matrix $\vec{A}_R^T \vec{A}_R + \epsilon \vec{I} $ for $100$ runs of the algorithm shown in orange for the set $R$ found---D-optimality maximizes the sum of eigenvalues for this submatrix. The eigenvalues are floored at $0.01$ because of regularization ($\epsilon = 0.01$). Best solution found has high sum of eigenvalues (brown dotted line)---planted D-optimal design has the maximal sum of eigenvalues (grey dashed line). Though the algorithms does not recover the planted design, it finds very competitive designs. meaning, high sum of eigenvalues of the Fisher Information Matrix. {\bf Right panel:} Count histogram of sum-log of eigenvalues of design found over the $100$ runs, compared to choosing random sets, both of size $R = 81$.}
    \label{fig:d-optimality-1}
\end{figure*}

In Fig.~\ref{fig:d-optimality-1} we show that the algorithm manages to find very competitive designs of size $81$, compared to a random selection of $81$ polymers. We run toe algorithm $100$ times. It does not recover the planted optimal design, however, the Fisher Information matrix of the best design found is full rank, and sum-log of eigenvalues of $\vec{A}^T_R \vec{A}_R$ (denoted by $\sum_i \log \lambda_i $ in Fig.~\ref{fig:d-optimality-1}) is close to the highest possible value---whereas a typical random sample is rank deficient. Note that in Fig.~\ref{fig:d-optimality-1}, eigenvalues are floored at $0.01$ because we chose the regularization constant $\epsilon = 0.01$ for the added regularization term ($\epsilon \vec{I}$) in computing the Fisher Information matrix eigenvalues.   

\section{Discussion on hyper-parameters and algorithmic efficiency}
\label{sec:hyper}
The approximate Integer Programming (AIP) algorithms applied to DoE in this work in Section~\ref{sec:AIP} and \ref{sec:DOE-IP} closely follows the work in Ref.~\cite{wu2018ell} on general Integer Programming. The successful application of ADMM to non-convex problems have been studied recently in the literature~\cite{wu2018ell, wang2019global, liu2019linearized, themelis2020douglas} including convergence analysis. We refer the reader to the literature for convergence analysis---our focus in this work is practical applications in polymeric DoE. We simplify the work in Ref.~\cite{wu2018ell} considerably. In our computational experiments, we observe that there are four important hyper-parameters in the algorithm after simplification from Ref.~\cite{wu2018ell}. 

They are $p$---the power of the $l_p$-constraint, $\rho_\textrm{init}$--- the initial value of the penalty parameter $\rho$, $\rho_\textrm{scaling}$---the scale factor that multiplies $\rho$ every $t_\textrm{step}$ steps to increase its value, thereby increasing the penalty stringency. We fix all of the penalty parameters to be equal,  $\rho_1 = \rho_2 = \rho_3$. We choose the power $p$ of the $l_p$ constraint to be $p= 3$. In Ref.~\cite{wu2018ell} it has been shown that for good convergence, $\rho$ needs to be initialized at a moderate value, our $\rho_\textrm{init} = 10$. The only hyper-parameters that need tinkering with are $\rho_{scaling}$ which we typically set at $1.05$, and the $t_\textrm{step}$ which we typically set to $10$ steps, but convergence can be sensitive to its value. In practice, hyper-parameters sweep was not needed for our test case problems.  

The algorithm presented in Section~\ref{sec:Sparse} has only one hyper-parameter, the sparsity parameter $\lambda$ and we set it at $\lambda = 0.9$. This has uniformly worked well in all test cases. 

The algorithm presented in Section~\ref{eq:DOE-IP} introduces a new hyper-parameter $t_\textrm{v-step}$ which controls how often we update $\vec{v}^k$ in Eq.~\ref{eq:conjugate_gradient2}. We choose this typically between $5-10$ steps. 

\section{Conclusions}
In this work, we present three new algorithms to solve challenging Design of Experiment problems (DoE) pertaining to design of polymers. To review, the general design setup we solve for is as follows: there exists an experimental setup to measure a set of target properties of copolymers that we want to engineer; our goal is to find the minimal set of polymer designs to measure in the experimental setup  in order to efficiently quantify the contributions to the polymer properties of the constituent positional-\hyperlink{kmer}{k-mers}; the contributions could be at individual-, pairwise-, etc. settings, or more broadly across arbitrary numerical features of these polymers. However, not all polymer designs are permissible owing to various experimental constraints---the problem is to identify an optimal subset picked from a much larger set of permissible polymers, in order to test them in parsimonious experimental designs, such that the dataset generated in this process is as unbiased as possible within the fixed experimental budget. These sort of problems are commonplace in a broad number of polymer engineering challenges, especially when experiments are costly to run, experimental and synthesis conditions introduce constraints etc. 

We argue that a large class of such problems can be mapped to finding densest subgraph of size $K$ in typically very dense graphs. We introduce and test two new algorithms to approximately solve the $K$-densest subgraph problem for such dense graphs. The two Integer Programming algorithms were inspired by Ref.~\cite{wu2018ell} that applied the ADMM method to integer programming. We show that both of our algorithms perform well by recovering planted cliques or finding near-optimal solutions that are much denser than random subgraphs. We believe that our algorithms provide a valuable, efficient, and pragmatic tool for finding approximate solutions of the NP-hard polymer DoE problems. We provide illustrative examples for such solution in design of polymeric molecules like nucleic acids. 

It has not escaped our attention that these algorithms have broader applicability for solving $K$-densest subgraph problems in settings where the underlying graph is quite dense---a regime not commonly addressed in the literature focused on real-world networks (social, web, communication etc.) which are not very dense on average and have only a few nodes of high degree (called \emph{hubs}). We hope future work will test performance of the algorithms presented herein against algorithms developed for such networks. Analysis of the time and memory cost of our algorithm, beyond what is already known for ADMM methods, is also left to future work. We observe that the sparse Conjugate Gradient (CG) method employed to solve a linear sub-problem at every step of ADMM is the most costly computational step in the algorithms presented here. In Section~\ref{sec:results} we have tested the algorithms on graphs of reasonably large size ($\sim 5000$ nodes).  

In the most general setting of polymer design, the polymer features could be simply numerical vectors. These vectors could include, for example, chemical descriptors or measurements of supplementary polymer properties informative in engineering the target properties of interest. In such scenarios, the design of a subset of polymers from a given set is not limited to just selecting \emph{balanced} \hyperlink{kmer}{k-mers} composition, but also \emph{balanced} feature values across the selected subset. For example, such a physical feature could be lipophilicity measurements of oligonucleotides in the target property of engineering biodistribution. For such cases, we present a new algorithm for identifying a subset of polymers from a larger set such that the subset is approximately D-optimal. This problem involves maximizing the determinant of a sub-matrix, and we test our algorithm on binary matrices for balanced designs of positional-pairwise k-mer compositions. Because the problem formulations are non-convex and therefore does not have global optima, we report distributions of solutions, and observe that good approximately D-optimal solutions are found. We note that the algorithm presented is broadly applicable to design of molecules beyond polymers, and in general, any experimental design that seeks approximate D-optimality.  

In summary, we believe that the algorithms presented here significantly advances the field of Design of Experiments paradigm to polymer applications, and more broadly to design of molecules for which compositional units are identifiable. We have used these algorithms in our work on engineering of safe and efficacious Oligonucleotide-based Medicines (OBMs).

\section{Acknowledgements}
The author acknowledges David Pekker and Jesse Levitt for useful discussions, refinement of the narrative and numerous editorial inputs; Sankha Pattanayak for illustrating Figure~\ref{fig:nucleic-acids}.

\bibliography{mybib}

\end{document}